\documentclass[showpacs,rotating,preprintnumbers,amsmath,amssymb]{elsart}
\usepackage{rotating}
\usepackage{graphicx}
\usepackage{dcolumn,color}
\usepackage{bm}
\usepackage{epsfig}
\begin{document}
\bibliographystyle{try}
\newcommand{\er}{$\pm$}
\newcommand{\be}{\begin{eqnarray}}
\newcommand{\ee}{\end{eqnarray}}
\newcommand{\widt}{\rm\Gamma_{tot}}
\newcommand{\wadd}{$\rm\mathcal B_{miss}$}
\newcommand{\gpiN}{$\rm\mathcal B_{\pi N}$}
\newcommand{\getN}{$\rm\mathcal B_{\eta N}$}
\newcommand{\gkla}{$\rm\mathcal B_{K \Lambda}$}
\newcommand{\gksi}{$\rm\mathcal B_{K \Sigma}$}
\newcommand{\gNpi}{$\rm\mathcal B_{P_{11} \pi}$}
\newcommand{\gDpf}{$\rm\mathcal B_{\Delta\pi(L\!<\!J)}$}
\newcommand{\gDps}{$\rm\mathcal B_{\Delta\pi(L\!>\!J)}$}
\newcommand{\sqgDpf}{$\rm\sqrt\mathcal B_{\Delta\pi(L\!<\!J)}$}
\newcommand{\sqgDps}{$\rm\sqrt\mathcal B_{\Delta\pi(L\!>\!J)}$}
\newcommand{\gnsi}{$\rm N\sigma$}
\newcommand{\gNpf}{$\rm\mathcal B_{D_{13}\pi}$}
\newcommand{\gNps}{$\mathcal B_{D_{13}\pi(L\!>\!J)}$}
\newcommand{\roper}{$ N(1440)P_{11}$}
\newcommand{\srma}{$  N(1535)S_{11}$}
\newcommand{\trma}{$ N(1520)D_{13}$}
\newcommand{\srmb}{$ N(1650)S_{11}$}
\newcommand{\trmb}{$ N(1700)D_{13}$}
\newcommand{\trmc}{$ N(1875)D_{13}$}
\newcommand{\trmd}{$ N(2170)D_{13}$}
\newcommand{\fvma}{$ N(1675)D_{15}$}
\newcommand{\fvmb}{$ N(2070)D_{15}$}
\newcommand{\fvpa}{$ N(1680)F_{15}$}
\newcommand{\srpb}{$ N(1710)P_{11}$}
\newcommand{\trpa}{$ N(1720)P_{13}$}
\newcommand{\trpb}{$ N(2200)P_{13}$}
\newcommand{\trpd}{$ N(2170)D_{13}$}
\newcommand{\dtpa}{$\Delta(1232)P_{33}$}
\newcommand{\doma}{$\Delta(1620)S_{31}$}
\newcommand{\dtma}{$\Delta(1700)D_{33}$}
\newcommand{\dtmb}{$\Delta(1940)D_{33}$}
\newcommand{\rthe}{A^{1/2}/A^{3/2}}
\newcommand{\amoh}{$A^{1/2}$}
\newcommand{\amth}{$A^{3/2}$}
\newcommand{\broh}{\Gamma^{1/2}_{\gamma p}/\widt}
\newcommand{\brth}{\Gamma^{3/2}_{\gamma p}/\widt}
\newcommand{\btot}{$\Gamma(\gamma p)$}
\newcommand{\Dpi}{\Delta\pi}
\newcommand{\KL}{\rm\Lambda K}
\newcommand{\KS}{\rm\Sigma K}

\newcounter{univ_counter}
\setcounter{univ_counter} {0}
\addtocounter{univ_counter} {1}
\edef\HISKP{$^{\arabic{univ_counter}}$ } \addtocounter{univ_counter}{1}
\edef\GIESSEN{$^{\arabic{univ_counter}}$ } \addtocounter{univ_counter}{1}
\edef\GATCHINA{$^{\arabic{univ_counter}}$ } \addtocounter{univ_counter}{1}
\edef\ERLANGEN{$^{\arabic{univ_counter}}$ } \addtocounter{univ_counter}{1}
\edef\FSU{$^{\arabic{univ_counter}}$ } \addtocounter{univ_counter}{1}
\edef\PI{$^{\arabic{univ_counter}}$ } \addtocounter{univ_counter}{1}
\edef\BOCHUM{$^{\arabic{univ_counter}}$ } \addtocounter{univ_counter}{1}
\edef\BASEL{$^{\arabic{univ_counter}}$ } \addtocounter{univ_counter}{1}
\edef\KVI{$^{\arabic{univ_counter}}$ } \addtocounter{univ_counter}{1}
\begin{frontmatter}

\title{\boldmath$N^*$ and \boldmath$\Delta^*$\unboldmath ~ decays into \boldmath$
N\pi^0\pi^0$\unboldmath }

\collab{The CB-ELSA Collaboration}
\author[HISKP,GIESSEN]{U.~Thoma},
\author[HISKP]{M.~Fuchs},
\author[HISKP,GATCHINA]{A.V.~Anisovich},
\author[ERLANGEN]{G.~Anton},
\author[PI]{R.~Bantes},
\author[HISKP]{O.~Bartholomy},
\author[HISKP]{R.~Beck},
\author[GATCHINA]{Yu.~Beloglazov},
\author[HISKP,FSU]{V.~Crede},
\author[HISKP]{A.~Ehmanns},
\author[HISKP]{J.~Ernst},
\author[HISKP]{I. Fabry},
\author[BOCHUM]{H.~Flemming},
\author[ERLANGEN]{A.~F\"osel},
\author[HISKP]{Chr.~Funke},
\author[PI]{R.~Gothe},
\author[GATCHINA]{A.~Gridnev},
\author[HISKP]{E.~Gutz},
\author[PI]{St.~H\"offgen},
\author[HISKP]{I.~Horn},
\author[ERLANGEN]{J.~H\"o\ss l},
\author[HISKP]{J.~Junkersfeld},
\author[HISKP]{H.~Kalinowsky},
\author[PI]{F.~Klein},
\author[HISKP]{E.~Klempt},
\author[BOCHUM]{H.~Koch},
\author[PI]{M.~Konrad},
\author[BOCHUM]{B.~Kopf},
\author[BASEL]{B.~Krusche},
\author[PI]{J.~Langheinrich},
\author[KVI]{H.~L\"ohner},
\author[GATCHINA]{I.~Lopatin},
\author[HISKP]{J.~Lotz},
\author[BOCHUM]{H.~Matth\"ay},
\author[PI]{D.~Menze},
\author[HISKP,GATCHINA]{V.A.~Nikonov},
\author[GATCHINA]{D.~Novinski},
\author[PI]{M.~Ostrick},
\author[HISKP]{H.~van~Pee},
\author[HISKP,GATCHINA]{A.V.~Sarantsev},
\author[HISKP]{C.~Schmidt},
\author[PI]{H.~Schmieden},
\author[PI]{B.~Schoch},
\author[ERLANGEN]{G.~Suft},
\author[GATCHINA]{V.~Sumachev},
\author[HISKP]{T.~Szczepanek},
\author[PI]{D.~Walther},
\author[HISKP]{Chr.~Weinheimer}\\

\address[HISKP]{Helmholtz-Institut f\"ur Strahlen- und Kernphysik der
Universit\"at Bonn, Germany}
\address[GIESSEN]{II. Physikalisches Institut, Universit\"at Giessen}
\address[GATCHINA]{Petersburg
Nuclear Physics Institute, Gatchina, Russia}
\address[ERLANGEN]{Physikalisches Institut,
Universit\"at Erlangen, Germany}
\address[PI]{Physikalisches Institut,
Universit\"at Bonn, Germany}
\address[FSU]{Department of Physics, Florida State University, USA}
\address[BOCHUM]{Physikalisches Institut, Universit\"at Bochum, Germany}
\address[BASEL]{Physikalisches Institut, Universit\"at
Basel, Switzerland}
\address[KVI]{KVI, Groningen, Netherlands}

\date{\today}


\begin{abstract}
Decays of baryon resonances in the second and the third resonance
region into $N\pi^0\pi^0$ are studied by photoproduction of two
neutral pions off protons. Partial decay widths of $N^*$ and
$\Delta$* resonances decaying into $\Delta(1232)\pi$, $
N(\pi\pi)_{S}$, \roper$\pi$, and \trma$\pi$ are determined in a
partial wave analysis of this data and of data from other reactions.
Several partial decay widths were not known before. Interesting
decay patterns are observed which are not even qualitatively
reproduced by quark model calculations. In the second resonance
region, decays into $\rm\Delta(1232)\pi$ dominate clearly. The $\rm
N(\pi\pi)_{S}$-wave provides a significant contribution to the cross
section, especially in the third resonance region. The $\rm
P_{13}$(1720) properties found here are at clear variance to PDG
values.  \vspace{5mm}   
\end{abstract}

\end{frontmatter}

The structure of baryons and their excitation spectrum is one of the
unsolved issues of strong interaction physics. The ground states and
the low--mass excitations evidence the decisive role of SU(3) symmetry
and suggest an interpretation of the spectrum in constituent quark
models \cite{Capstick:bm,Glozman:1997ag,Loring:2001kx}. Baryon decays
can be calculated in quark models using harmonic--oscillator wave
functions and assuming a $q\bar q$ pair creation operator for meson
production. A collective string-like model gives a description of the
mass spectrum of similar quality \cite{Bijker:1994yr} and predicts
partial decay widths of resonances \cite{Bijker:1996tr}. A
comprehensive review of predictions of baryon masses and decays can be
found in~\cite{Capstick:2000qj}. An alternative description of the
baryon spectrum may be developed in effective field theories in which
baryon resonances are generated dynamically from their decays
\cite{Oset:2004dd}. At present, the approach is restricted to
resonances coupling to octet baryons and pseudoscalar mesons, yet it
can possibly be extended to include vector mesons and decuplet baryons
\cite{Lutz:2005ip}. To test the different approaches, detailed
information on the spectrum and decays of resonances is needed,
including more complex decay modes such as $\Delta\pi$ or
$N(\pi\pi)_{\rm S}$, where $(\pi\pi)_{\rm S}$ stands for the $(\pi\pi)$-$S$-wave. The
analysis of complex final states requires the use of event-based
likelihood fits to fully exploit the sensitivity of the data.
In baryon spectroscopy such fits have, to our knowledge, never been
performed so far.

In this letter we report on a study of $\Delta\pi$ and other
$p2\pi^0$ decay modes of baryon resonances belonging to the second and
third resonance region. The results are obtained from data on the
reaction
\begin{equation} \gamma p \to p \pi^0\pi^0\,. \label{2pi0}
\end{equation}
The data were obtained using the tagged photon beam of the {\bf
EL}ectron {\bf S}tretcher {\bf A}ccelerator (ELSA)
\cite{Hillert:2006yb} at the University of Bonn, and the Crystal
Barrel detector \cite{Aker:1992ny}. A short description of the
experiment, data reconstruction and analysis methods can be found in
two letters on single $\pi^0$ \cite{Bartholomy:2004uz} and $\eta$
\cite{Crede:2003ax} photoproduction, a more comprehensive one in
\cite{van Pee:2007tw,Bartholomy:2007}. The analysis presented here
differs only in the final state consisting now of four photons
(instead of two or six) and a proton.  The data cover the photon
energy range from 0.4 to 1.3\,GeV.

\begin{figure}[pb]
\begin{minipage}[c]{0.46\textwidth}
\epsfig{file=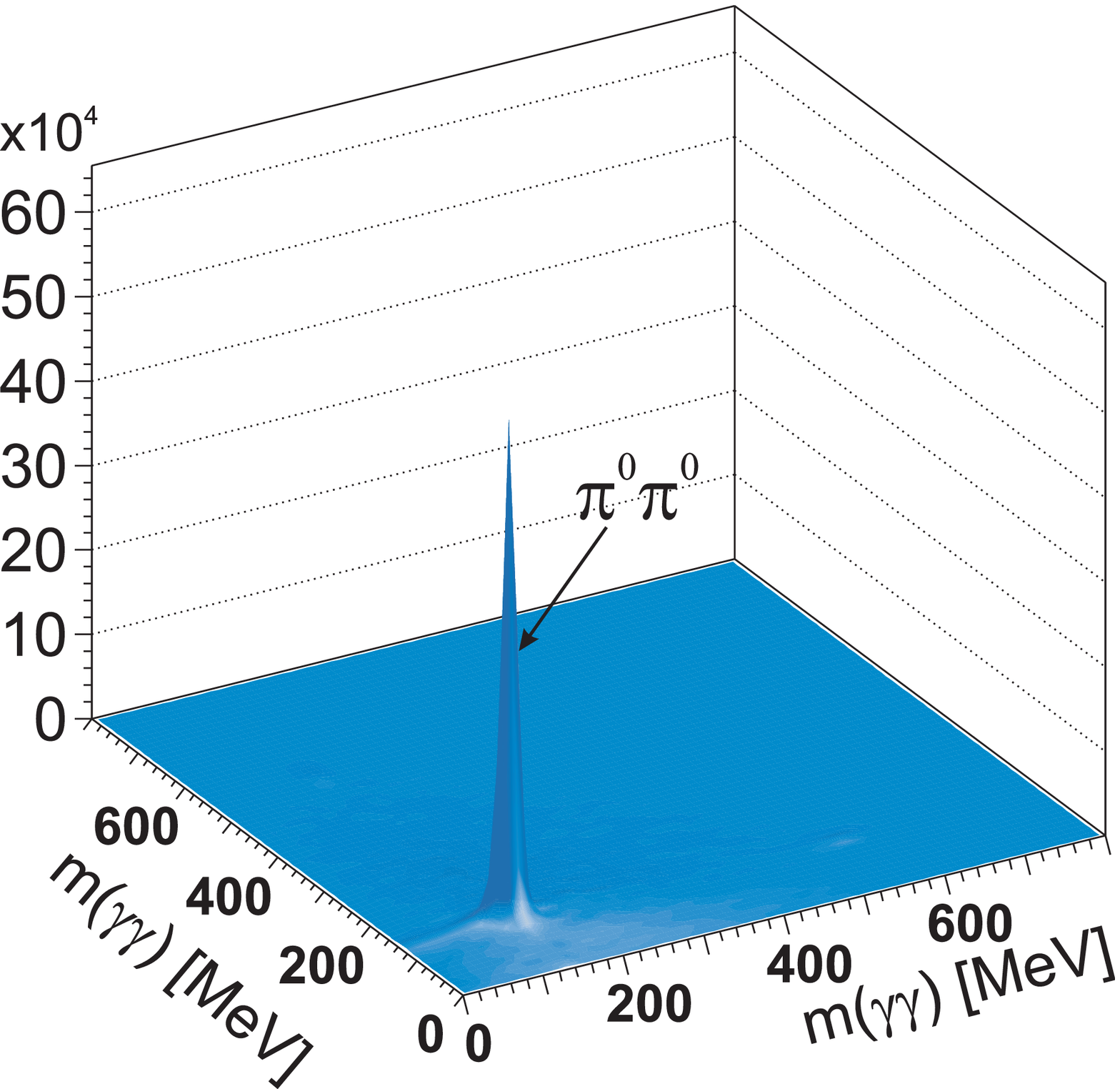,width=\textwidth,height=6cm,clip=}
\end{minipage}
\begin{minipage}[c]{0.48\textwidth}
\hspace{-0mm}\epsfig{file=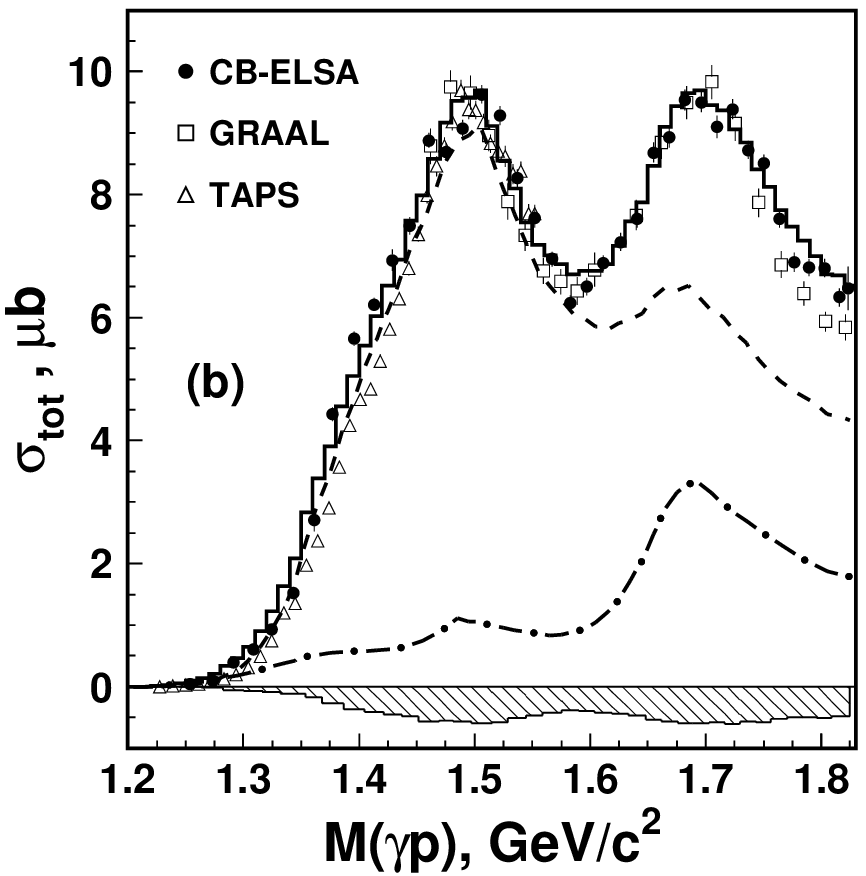,width=1\textwidth,height=6.6cm,clip=}
\end{minipage}
\caption{\label{gg_vs_gg} a) $\gamma\gamma$ invariant mass
distribution of one photon pair versus the invariant mass
distribution of the second photon pair after a kinematical fit to
$\gamma p \to p 4\gamma$ (6~entries per event). b) Total cross
sections for $  \gamma p \to p \pi^0\pi^0$, $\bullet$ this work;
${\tiny\triangle}$: TAPS~~\cite{Sarantsev:2007};
open squares: GRAAL~\cite{Assafiri:2003mv}. Solid line: PWA fit,
band below the figure: systematic error (see text). Dashed curve:
$ \Delta^+\pi^0\to p\pi^0\pi^0$, dashed-dotted line: $
p(\pi^0\pi^0)_{S}$ cross section as derived from the PWA.}
\end{figure}

In the analysis, events due to reaction (1) are selected by requiring
five clusters of energy deposits in the Crystal Barrel calorimeter, one
of them matching the direction of a charged particle emerging from the
liquid H$_2$ target of 5cm length and hitting a three--layer
scintillation fiber (Scifi) detector surrounding the target. The latter
cluster is assigned to be a `proton', the other four clusters are
treated as photons. Events are also retained when they have four
clusters in the calorimeter and a hit in the Scifi which cannot be
matched to any of the clusters. The Scifi hit is then treated as
`proton', the four hits in the barrel as photons. In a second selection
step, the events are subjected to a one--constraint kinematical fit to
the $\gamma p\to p4\gamma$ hypothesis imposing energy and momentum
conservation and assuming that the interaction took place in the target
center. The proton is treated as missing particle, its direction
resulting from the fit has to agree with the direction of the detected
proton  within 20$^{\circ}$. The $\gamma\gamma$ invariant mass
distribution of one photon pair versus the invariant mass distribution
of the second photon pair is plotted in Fig.~\ref{gg_vs_gg}a.
A $2\sigma$ cut ($\sigma=8$\,MeV/c$^2$) was applied to
the two $\pi^0$, then the mass of the two $\pi^0$ was imposed in a
$\gamma p \to p\pi^0\pi^0\to p\,4\gamma$ three--constraint
kinematical fit with a missing proton. Its confidence level had to
exceed 10$\%$ and had to be larger than that for a fit to $\gamma
p\to p\pi^0\eta\to p\,4\gamma$. The final event sample
contains~115.600 events. Performing extensive GEANT--based Monte
Carlo (MC) simulations, the background was shown to be less than
1$\%$. The acceptance determined from MC simulations vanishes for
forward protons leaving the Crystal Barrel through the forward hole,
and for protons going backward in the center-of-mass system, having
very low laboratory momenta. The overall acceptance depends on the
contributing physics amplitudes which are determined by a partial
wave analysis (PWA) described below. MC events distributed according
to the PWA solution were used to determine the correct acceptance.
This MC data sample undergoes the same analysis chain as real data.

We first discuss the main features of the data. Fig.~\ref{gg_vs_gg}b
shows the total cross section for $2\pi^0$ photoproduction together
with the $\Delta\pi$ and $p(\pi\pi)_{S}$ excitation functions. Two
peaks due to the second and third resonance region are immediately
identified. Our data points are given by black dots, the bars
re\-present the statistical errors. The systematic error due to the
acceptance correction is determined by the spread of results
obtained from different PWA solutions. A second systematic error is
due to uncertainties in the reconstruction~\cite{van Pee:2007tw}.
These errors are added quadratically to determine the total
systematic error shown as a band below the cross section. This error
does not contain the normalization uncertainty of $\pm
5\%$~\cite{van Pee:2007tw}.

The general consistency between our data and those from A2-TAPS
\cite{Sarantsev:2007} (superseding in statistics earlier MAMI data
\cite{Harter:1997jq,Wolf:2000qt}) and GRAAL \cite{Assafiri:2003mv}
is good (see Fig.~\ref{gg_vs_gg}b). In the low--energy region, our
data show a shoulder which is less pronounced in the A2-TAPS data
(see \cite{Sarantsev:2007}). The recent A2-GDH measurements
\cite{Ahrens:2005ia} fall in between these two results. The DAPHNE
data exceed our cross section significantly \cite{Braghieri:1994rf}.
At larger energies, the GRAAL data fall off with energy faster than
our data. Data taken at higher energies covering the photon energy
range from 0.8 to 3\,GeV yield a cross section~\cite{Fuchs} which is
compatible in the overlap region with the results presented here.
All 3 experiments do not cover the full solid angle. In this
analysis and in the analysis of the A2-TAPS collaboration, the cross
section is extrapolated into ``blind" detector regions using the
result of the partial wave analysis.  The GRAAL collaboration
simulates $\gamma p\to \Delta^+\pi^0$ and $\gamma p\to p\pi^0\pi^0$
to account for the acceptance.

\begin{figure}[pt]
\begin{center}
\begin{tabular}{ccc}
\epsfig{file=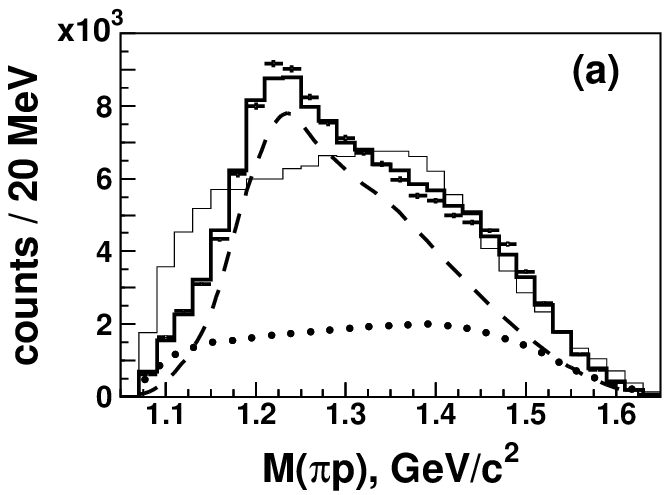,width=0.32\textwidth,height=0.17\textheight,clip=}&
\epsfig{file=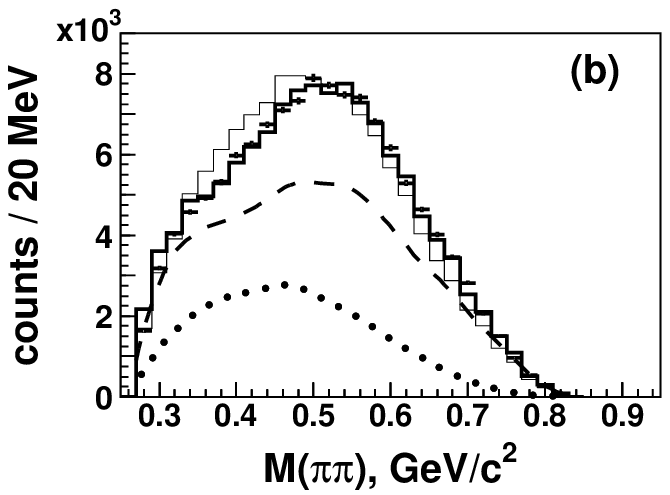,width=0.32\textwidth,height=0.17\textheight,clip=}&
\epsfig{file=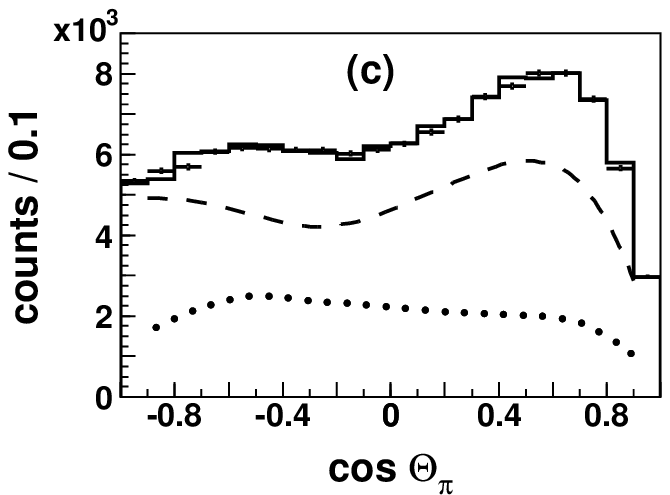,width=0.32\textwidth,height=0.17\textheight,clip=}\\
\epsfig{file=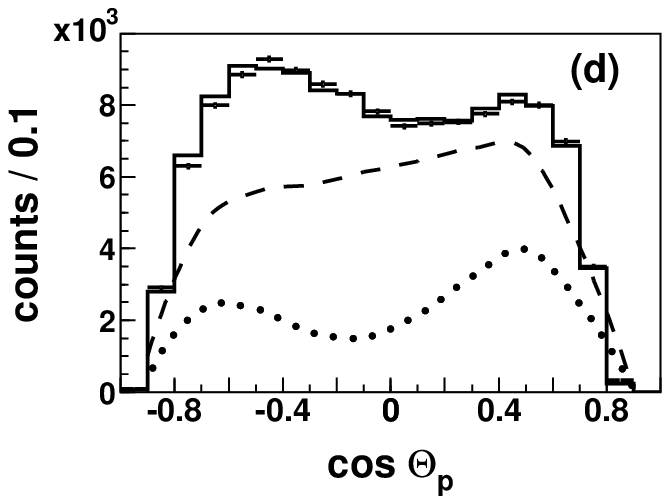,width=0.32\textwidth,height=0.17\textheight,clip=}&
\epsfig{file=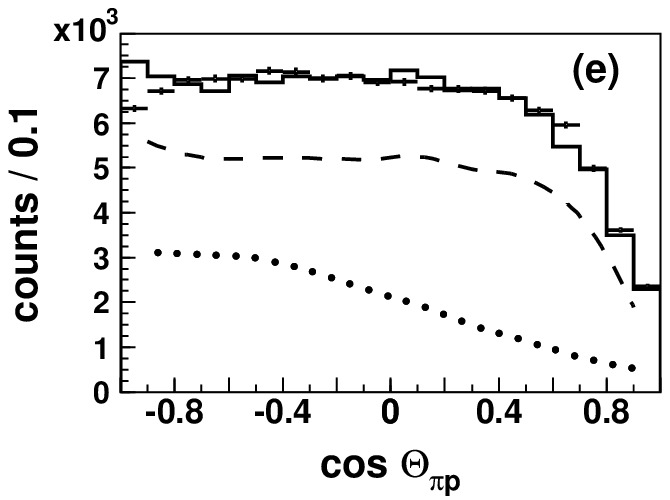,width=0.32\textwidth,height=0.17\textheight,clip=}&
\epsfig{file=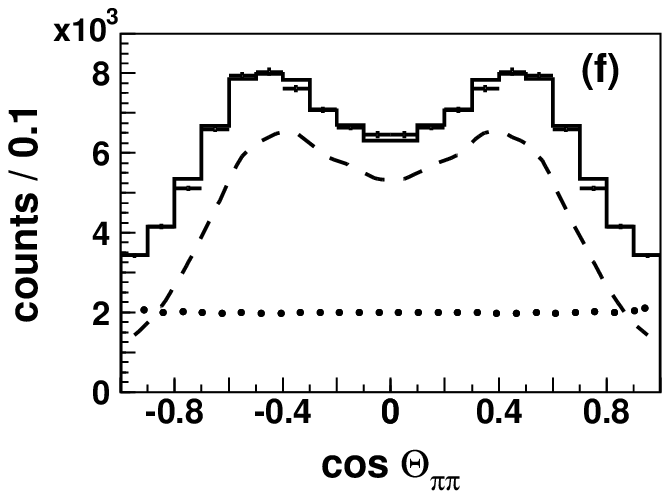,width=0.32\textwidth,height=0.17\textheight,clip=}\\
\end{tabular}
\end{center}
\caption{\label{dp} Mass and angular distributions for $\gamma p \to
p \pi^0\pi^0$ after a 1550--1800\,MeV/c$^2$ cut on $M_{
p\pi^0\pi^0}$. a: $ p\pi^0$, b: $\pi^0\pi^0$ invariant mass. In
c)--f) $\cos\theta$ distributions are shown. In (c), $\theta$ is the
angle of a $\pi^0$ with respect to the incoming photon in the
center--of--mass--system (cms); in (d), the cms angle of the proton
with respect to the photon is shown; in (e), the angle between two
pions in the $\pi^0 p$ rest frame; in (f) the angle between $\pi^0$
and $p$ in the $\pi^0\pi^0$ rest frame. Data are represented by
crosses, the fit as solid line, the thin line in a,b) shows the
phase space distribution. Dashed: $\Delta^+\pi^0\to p\pi^0\pi^0$,
dotted: $p(\pi^0\pi^0)_{S}$ contribution. The distributions are not
corrected for acceptance to allow a fair comparison of the fit with
the data without introducing any model dependence by extrapolating
e.g. over acceptance holes. Differential cross sections will be
given elsewhere. } \end{figure}

Fig.~\ref{dp}a,b) shows the $ p\pi^0$ and $\pi^0\pi^0$ invariant
mass distribution for reaction (1) after a 1550--1800\,MeV/c$^2$ cut in
the $ p\pi^0\pi^0$ mass. Also shown are some angular distributions.
The data and their errors are represented by crosses, the lines give
the result of the fits described below. The $ p\pi^0$ mass
distribution reveals the role of the $\Delta$ as contributing
isobar. The $\pi^0\pi^0$ mass distribution does not show any
significant structure. While $2\pi$ decays of resonances belonging to
the 2$^{\rm nd}$ resonance region are completely dominated by the
$\Delta\pi$ isobar as intermediate state, the two-pion S-wave provides a
significant decay fraction in the 3$^{\rm rd}$ resonance region.

\begin{figure}[pt]
\begin{center}
\epsfig{file=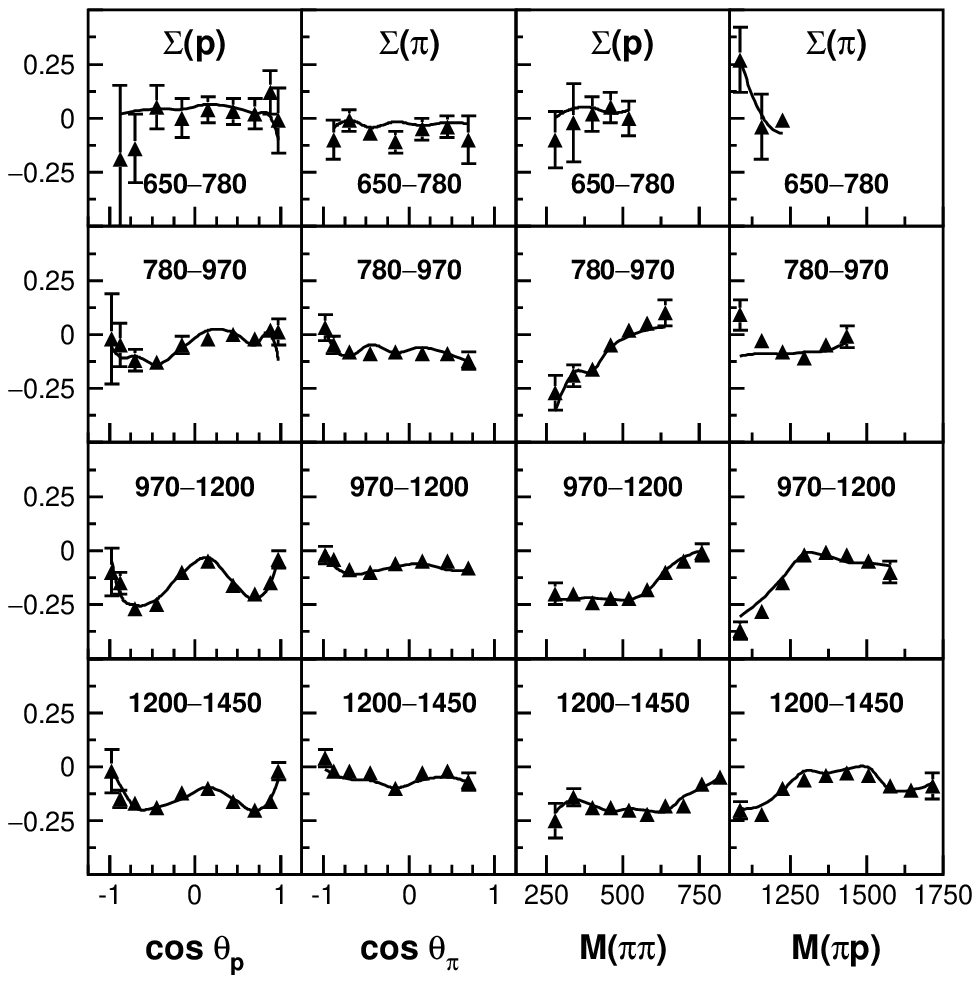,width=.90\textwidth,height=0.4\textheight}
\end{center}
\caption{\label{graal_pipi}The beam asymmetry $\Sigma$ for the reaction
$\gamma p\to p\pi^0\pi^0$ as a function of the proton or $\pi^0$
direction with respect to the beam axis, and as a function of the
$\pi^0\pi^0$  and $ p\pi^0$ invariant mass \cite{Assafiri:2003mv}.
The solid line represents the PWA fit. The numbers given in the
figures indicate the photon energy bin.
}\vspace{3mm} 
\begin{center}
\epsfig{file=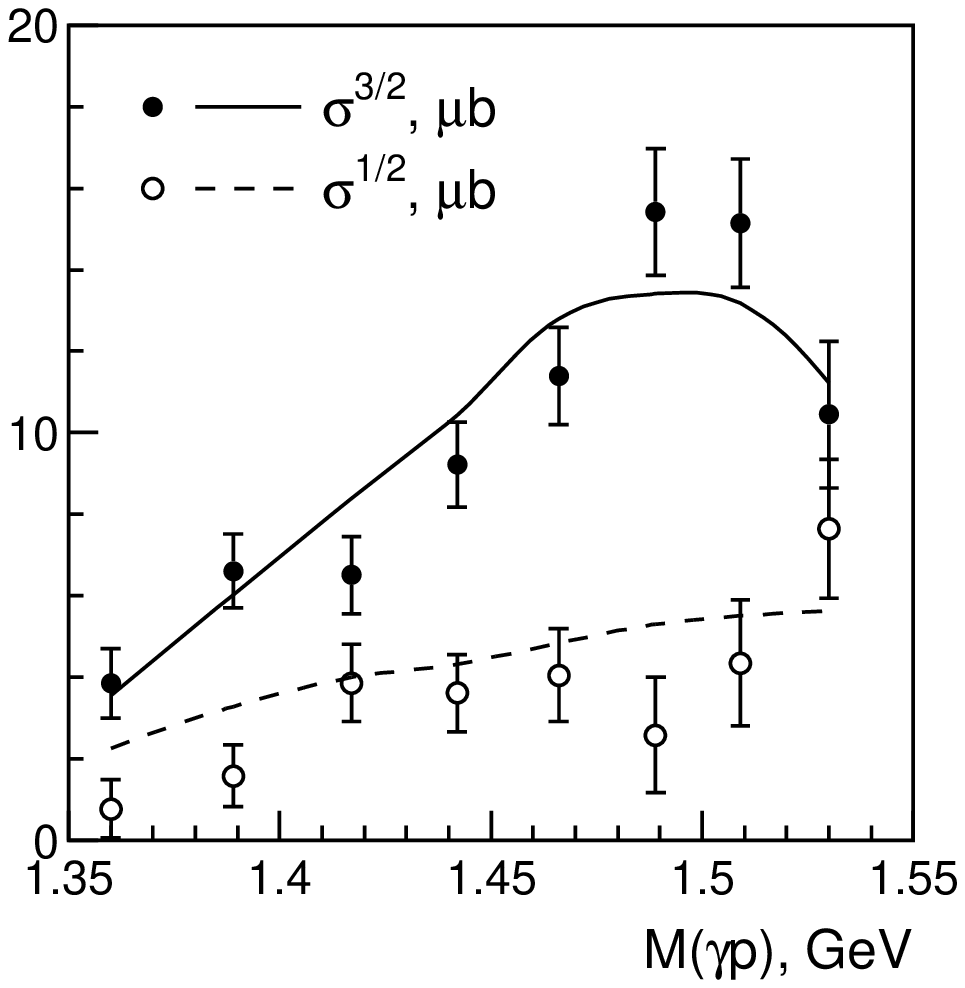,width=.70\textwidth,height=0.50\textwidth,height=0.25\textheight}
\end{center}
\caption{\label{mz_pipi}Helicity dependence of the reaction $\gamma p
\to p \pi^0\pi^0$ \cite{Ahrens:2005ia}. The lines represent the result
of the PWA fit.
 }
\end{figure}

The partial wave analysis uses an event--based maximum likelihood
fit.
 To constrain the analysis, not only
the data on reaction (1) were used in the fit but also data on
$\gamma p\to p\pi^0$
\cite{Bartholomy:2004uz,SAID1_1,SAID1_2,SAID1_3,SAID2_1,SAID2_2,SAID2_3,Bartalini:2005wx}
including differential cross sections, beam and target asymmetry,
and recoil polarization, further data on $\gamma p\to p\pi^0\pi^0$
\cite{Assafiri:2003mv,Ahrens:2005ia}, $\gamma p\to p\eta$
\cite{Crede:2003ax,Krusche:nv,GRAAL1,Bartalini:2007fg}, and data on
$\gamma p\to K\Lambda$, and $ K\Sigma$ \cite{Glander:2003jw,%
McNabb:2003nf,Zegers:2003ux,Lawall:2005np,Bradford:2005pt,%
Bradford:2006ba,Lleres:2007tx}.  The SAID $\pi N$ partial-wave
elastic scattering amplitudes \cite{Arndt:2006bf} are used to
constrain the K-matrices for the $S_{11}$, $P_{11}$, $P_{13}$,
$P_{33}$, $D_{33}$ partial waves. Details of the fitting procedure
and on the $\chi^2$ contributions of the different reactions are
given in \cite{Anisovich:2007}. As examples, we show in Fig.
\ref{graal_pipi} the beam asymmetry $\Sigma$ \cite{Assafiri:2003mv}
and in Fig. \ref{mz_pipi} the helicity dependence of the reaction
$\gamma p \to p \pi^0\pi^0$ \cite{Ahrens:2005ia}. Inclusion of the
beam asymmetry had an impact on the size of couplings but did not
lead to significant changes of the pole positions. The helicity
dependence was correctly predicted; correspondingly, its inclusion
had no effect on the final solution.

Particularly useful were the Crystal Ball data on the charge exchange
reaction $\pi^-p\to n\pi^0\pi^0$ \cite{Prakhov:2004zv}. Even though
limited to masses below 1.525\,GeV/c$^2$, the data provided also
valuable constraints for the third resonance region due to their long
low--energy tails. The log likelihoods of the different data sets are
added with some weights varying from 1 to up to 30
\cite{Anisovich:2007}. The weights are chosen to force the fits to
describe low-statistics data with reasonable accuracy even on the
expense of a worse description of high-statistics data, where large
$\chi^2$ contributions can be the result of small deficiencies due to
model imperfections. The data on $2\pi^0$ photoproduction enter with a
weight 4. Moderate changes in the weights lead to changes in the
results which are covered by the quoted errors.

We started the analysis from the solution given in
\cite{Anisovich:2005tf,Sarantsev:2005tg} and found good compatibility.
The new $p\pi^0\pi^0$ data provides information on the N$\pi\pi$ decay
modes, without inducing the need to change masses or widths of the
contributing resonances (from \cite{Anisovich:2005tf,Sarantsev:2005tg})
beyond their respective errors, even though all parameters were allowed
to adjust again. The quality of the fits of the previous data did not
worsen significantly due to the constraints by the new $p\pi^0\pi^0$-data.

The dynamical amplitudes comprise resonances and background terms
due to Born graphs and $t$-- and $u$--channel exchanges. Angular
distributions are calculated using relativistic operators
\cite{Anisovich:2004zz}. Relations between cross sections and
resonance partial widths are given in \cite{Anisovich:2006}. Most
partial waves are described by multi--channel Breit-Wigner
amplitudes with an energy dependent width (in the form suggested by
Flatt\'e \cite{Flatte:1976xu}). Partial widths are calculated at the
position of the Breit-Wigner mass. For the K-matrix
parameterizations the Breit-Wigner parameters are determined in the
following way. First, the couplings are calculated as T-matrix pole
residues, then the imaginary part of the Breit-Wigner denominator is
parameterized as a sum of these couplings squared, multiplied by the
corresponding phase volumes and scaled by a common factor. This
factor as well as the Breit-Wigner mass are chosen as to reproduce
the amplitude pole position on the Rieman sheet closest to the
physical region. The Breit-Wigner parameters of the $
S_{11}$-resonances are determined without taking into account the
$\Delta\pi$-width to obtain results which can be compared with the
Particle Data Group (PDG) values \cite{Eidelman:2004wy}.

Table~\ref{resonance_list} summarizes the results of our fits. In
the absence of double-polarization data, there is no unique
solution. We have studied a large variety of solutions and estimated
the errors in the Table from the range of values found for different
solutions giving an acceptable description of the data. Most results
agree, within their respective errors, reasonably well with previous
findings. The errors quoted are estimated from the variance of
results of a large number of fits which provide an adequate
description of the data. Several partial decay widths for baryon
decays into $N\pi\pi$ were not known before. For widths known from
previous analyses, good compatibility is found. The helicity
amplitudes quoted in the table are calculated at the position of the
resonance pole. Hence they acquire a phase. As long as the phase is
small, the comparison with PDG values is still meaningful. We now
discuss a few partial waves.

{\bf The \boldmath$P_{13}$ wave} is described by a three-pole
multi-channel K-matrix which we interpret as $N(1720)P_{13}$,
$N(1900)P_{13}$, and $N(2200)P_{13}$. The $N(1900)P_{13}$ resonance
is required \cite{Nikonov:2007} due to the inclusion of the CLAS
spin transfer measurements in hyperon photoproduction
\cite{Bradford:2006ba}. The $N(2200)P_{13}$ was already needed to
fit single-pion photoproduction \cite{Anisovich:2005tf}.

\begin{table}[pb]
\caption{\label{resonance_list} Properties of the resonances
contributing to the $\gamma p\to \pi^0\pi^0 p$ cross section. The
masses and widths are given in MeV, the branching ratios $\mathcal
B$ in \% and helicity couplings in GeV$^{-1/2}$. The helicity
couplings and phases were calculated as residues in the pole
position which is denoted as `Mass' and `$\Gamma_{tot}$'. The method
for calculation of Breit-Wigner parameters is described in the
text.}
\end{table}
\renewcommand{\arraystretch}{1.8}                                     %
\begin{sideways}
\scriptsize \begin{tabular}{lccccccccc} \hline\hline
&\srma&\srmb&\trma&\trmb&\fvma&\trpa&\fvpa&\doma&\dtma        \\
\hline                                                                %
Mass&1508$^{+10}_{-30}$&1645\er15&
1509\er7&1710\er15&1639\er10&1630\er90&1674\er5&1615\er25&
1610\er35 \vspace*{-2mm} \\
{\tiny\phantom{rrrrr}\hfill PDG }&
{\tiny1495--1515}&{\tiny1640--1680}&{\tiny1505--1515}&{\tiny1630--1730}&
{\tiny1655--1665}&{\tiny1660--1690}&{\tiny1665--1675}&{\tiny1580--1620}&{\tiny1620--1700}\vspace*{-0mm}\\
$\widt$ &165\er15&187\er20&
 113\er12& 155\er25&
 180\er20&460\er80&95\er10&{180\er35} & 320\er60 \vspace*{-2mm}\\
{\tiny\phantom{rrrrr}\hfill PDG }& {\tiny90--250}
&{\tiny150--170}&{\tiny110--120}&{\tiny50--150}&{\tiny125--155}&{\tiny115--275}&
{\tiny105--135}&{\tiny100--130}&{\tiny150--250}\vspace*{-0mm} \\
\hline                                                                %
M$_{BW}$& 1548\er15& 1655\er15& 1520\er10&
 1740\er20& 1678\er15&1790\er100&1684\er8&
 1650\er25&1770\er 40\vspace*{-2mm} \\
{\tiny\phantom{rrrrr}\hfill PDG  }&
{\tiny1520--1555}&{\tiny1640--1680}&{\tiny1515--1530}&{\tiny1650--1750}&
{\tiny1670--1685}&{\tiny1700--1750}&{\tiny1675--1690}&{\tiny1615--1675}&{\tiny1670--1770}\vspace*{-0mm} \\
$\Gamma^{BW}_{tot}$ & 170\er20&180\er20& 125\er15&
 180\er30& 220\er25& 690\er100& 105\er8&
250\er60& 630\er150 \vspace*{-2mm} \\
{\tiny\phantom{rrrrr}\hfill PDG }&
{\tiny100--200}&{\tiny145--190}&{\tiny110--135}&{\tiny50--150}&{\tiny140--180}&
{\tiny150--300}&{\tiny120--140}&{\tiny120--180}&{\tiny200--400}\vspace*{-0mm} \\
\hline $A_{1/2}$ & 0.086\er0.025& 0.095\er0.025& 0.007\er0.015&
0.020\er0.016& 0.025\er0.01& 0.15\er0.08&-(0.012\er0.008)&
0.13\er0.05&  0.125\er0.030 \vspace*{-2mm} \\
{\tiny\phantom{rrrrr}\hfill phase  }&
{\tiny$(20\pm15)^{\circ}$}&{\tiny$( 25\pm20)^{\circ}$}&
{\tiny$-$}&{\tiny$-(4\pm5)^{\circ}$}&
{\tiny$-(7\pm5)^{\circ}$}&{\tiny$-(0\pm25)^{\circ}$}&
{\tiny$-(40\pm15)^{\circ}$}&{\tiny$-(8\pm5)^{\circ}$}&
{\tiny$-(15\pm10)^{\circ}$}\vspace*{-2mm} \\
{\tiny\phantom{rrrrr}\hfill PDG }&
{\tiny$(0.090\pm0.030)$}&{\tiny$(0.053\pm0.016)$}&
{\tiny-$(0.024\pm0.009)$}&{\tiny$-(0.018\pm0.013)$}&
{\tiny0.019\er0.008} &{\tiny0.018\er0.030}&
{\tiny$-(0.015\pm0.006)$}&{\tiny$0.027\pm0.011$}&
{\tiny$(0.104\pm0.015)$}\\
$A_{3/2}$&&& 0.137\er0.012&0.075\er0.030& 0.044\er0.012& 0.12\er0.08&
 0.120\er0.015&&  0.150\er0.060\vspace*{-2mm}\\
{\tiny\phantom{rrrrr}\hfill phase}&&&
{\tiny$-(5\pm5)^{\circ}$}&{\tiny$-(6\pm8)^{\circ}$}&{\tiny$-(7\pm5)^{\circ}$}&{\tiny$-(20\pm40)^{\circ}$}&
{\tiny$-(5\pm5)^{\circ}$}&& {\tiny$-(15\pm10)^{\circ}$}\vspace*{-2mm}\\
{\tiny\phantom{rrrrr}\hfill PDG}&&&{\tiny$0.166\pm0.005$}&{\tiny$-(0.002\pm0.024)$}&
{\tiny0.015\er0.009}  &{\tiny-(0.019\er0.020)}&{\tiny0.133\er0.012}
&&{\tiny 0.085\er0.022} \vspace*{-0mm} \\
\hline
\wadd &-&-&13\er5\,\%&20\er15\,\%&20\er8\,\%&-&2\er2\,\% &
10\er7\,\%&15\er10\,\%\vspace*{-2mm}\\
{\tiny\phantom{rrrrr}PDG($N\rho$) }& {\tiny$<4$\,\%}
&{\tiny4--12\,\%}&{\tiny15--25\,\%}&{\tiny$<35$\,\%}&
{\tiny$<$1--3\,\%}&{\tiny70--85\,\%}&{\tiny3--15\,\%}&{\tiny7--25\,\%}&{\tiny30--55}\vspace*{-0mm}\\
\gpiN &37\er9\,\%&
70\er15\,\%& 58\er8\,\%&8$^{+8}_{-4}$\,\%&
 30\er8\,\%&9\er6\,\%&72\er15\,\%& 22\er12\,\%&15\er8\,\%
\vspace*{-2mm} \\ {\tiny\phantom{rrrrr}\hfill PDG }&
{\tiny35--55\,\%}&{\tiny55--90\,\%}&{\tiny50--60\,\%}&{\tiny5--15\,\%}&
{\tiny40--50\,\%}&{\tiny10--20\,\%}&{\tiny60--70\,\%}&{\tiny10--30\,\%}&{\tiny10--20\,\%}\vspace*{-0mm} \\
\getN &40\er10\,\%& 15\er6\,\%&0.2\er0.1\,\%&
 10\er5\,\%&  3\er 3\,\% &10\er7\,\%&$<1$\,\% & - & - \vspace*{-1mm}\\
{\tiny\phantom{rrrrr}\hfill PDG\  }&
{\tiny30--55\,\%}&{\tiny3--10\,\%}&{\tiny0.23\er0.04\,\%}&{\tiny0\er1\,\%}&
{\tiny0\er1\,\% }&{\tiny4\er1\,\%}& {\tiny0\er1\,\%}&& \vspace*{-0mm}\\
\gnsi &-&-&$<4$\,\%& 18\er12\,\% & 10\er5 &3\er3\,\% &  11\er5\,\% &
- &  \vspace*{-2mm} \\
{\tiny\phantom{rrrrr}\hfill PDG\ }&&{\tiny$<4$\,\%}
&{\tiny$<8$\,\%}&&&-&{\tiny 5--20\,\%}&&\vspace*{-0mm}\\
\hline \gkla
&-& 5\er5\,\%&-&1\er1\,\%& 3\er3\,\% & 12\er9\,\% & $<1$\,\% & &
-\vspace*{-1mm}\\ \gksi
&-&-&-&$<1$\,\%&$<1$\,\%&$<1$\,\%&$<1$\,\%\vspace*{-0mm}\\
\hline \gDpf & & &12\er4\,\%& 10\er5\,\%
&24\er8\,\%&38\er20\,\%&8\er3\,\% & 48\er25\,\% &\vspace*{-2mm}\\
{\tiny$L<J$\quad\ PDG\ }&
&&{\tiny5--12\,\%}&&&-&
{\tiny6--14\,\%}&{\tiny30--60\,\%}&\vspace*{-0mm}\\[-4ex]
&&&&&&&&& 70\er20\,\%\vspace*{-2mm}\\
&&&&&&&&&{{\tiny30--60\,\%}}\vspace*{-0mm}\\[-4ex]
\gDps &23\er8\,\% &10\er5\,\% & 14\er5\,\%& 20\er11\,\% &$<3$\,\%
&7\er8\,\%& 4\er3\,\% &&  \vspace*{-2mm}\\
{\tiny$L>J$\quad\ PDG\ }&{\tiny$<$1\,\%}&&
{\tiny10-14\,\%}&&&-&{\tiny$<2$\,\%}&{\tiny} & \\
\hline \gNpi
&&&2\er2\,\%&14\er8\,\% &$<3$\,\%   &-& -  & 19\er12\,\%  & $<5$\,\%
\\ \gNpf &&& -               & -         &  4\er4\,\%  &24\er20\,\%& -    & - &  $<3$\,\%
\\ \hline \hline
\end{tabular}
\end{sideways}
\renewcommand{\arraystretch}{1.0}

Here, only the $N(1720)P_{13}$ resonance is discussed; for further
information, see \cite{Nikonov:2007}.  The $N(1720)P_{13}$ resonance
is the only resonance with properties which are clearly at variance
with PDG values. The central value for its total width
$\Gamma_{tot}^{BW}$ is 400 to 500\,MeV compared to the 200\,MeV
estimate of the PDG. However, Manley {\it et al.}
\cite{Manley:1992yb} find $(380\pm 180)$\,MeV/c$^2$. Its strongest
decay mode is found to be  $\Delta\pi$, not reported
in~\cite{Eidelman:2004wy}.  We find a rather small missing width of
(6\er 1)\% of the total width while the PDG assigns $70-85\%$ to the
$N\rho$-decay mode. A similar discrepancy was observed in
electro-production of two charged pions \cite{Ripani:2002ss}, and
interpreted either as evidence for a new -- rather narrow -- $
P_{13}$-state or as a wrong PDG $N\rho$-decay width. In agreement
with \cite{Aznauryan:2003zg}, we find a large branching ratio for
\trpa$\to N\eta$ while most analyses ascribe the $ N\eta$ intensity
in this mass region to \srpb.

{\bf The \boldmath$ P_{33}$ wave} is represented by a two-pole
two-channel K-matrix. The low energy part of pion photoproduction is
described by the $\Delta(1232)$ state even though non-resonant
contributions were needed to get a good fit. The quality of the
description of the elastic amplitude improved dramatically by
introduction of a second pole. The first K-matrix pole has $1231\pm
4$\,MeV/c$^2$ mass and helicity couplings $a_{1/2}=-0.125\pm 0.008$
and $a_{3/2}=-0.267\pm 0.010$. The pole position in the complex
energy plane was found to be $M=1205\pm 4$\,MeV/c$^2$ and $2\times
Im=92\pm 10$\,MeV/c$^2$.  The second K-matrix pole was not very
stable and varied between 1650 and 1800\,MeV/c$^2$. The T-matrix
pole showed better stability, and gave $M=1550\pm 40$\,MeV/c$^2$ and
$\Gamma=290\pm 60$\,MeV/c$^2$. This can be compared to the PDG
ranges, $M=1550-1700$\,MeV/c$^2$ and $\Gamma=250-450$\,MeV/c$^2$.

{\bf The two \boldmath$ S_{11}$ resonances}
(Table~\ref{resonance_list}) are treated as coupled--channel
$5\otimes 5$ K-matrix including $ N\pi$, $ N\eta$, $ K\Lambda$, $
K\Sigma$, and $\Delta\pi$ as channels. The $ N\sigma$ or the $
N\rho$ decay mode were added as 6$^{\rm th}$ channel for part of the
fits. The first K-matrix pole varied over a wide range in different
fits, from 1100 to 1480\,MeV/c$^2$. The physical amplitude
(T-matrix) exhibited, however, a stable pole at
M$_{pole}$=1508$^{+10}_{-30}$\,-i(83$\pm$8)\,MeV/c$^2$, in good
agreement with PDG. This pole position is very close to the $\eta N$
threshold. In some fits the pole moved under the $\eta N$ cut; in
that case the closest physical region for this pole is the $\eta N$
threshold. No other pole around 1500\,MeV/c$^2$ close to the
physical region was then found on any other sheet.  The second
K-matrix pole always converged to $1715\pm 30$ MeV/c$^2$ resulting
in a T-matrix pole as given in Table~\ref{resonance_list}.
Introduction of an additional pole did not lead to a significant
improvement in the fit.

{\bf The \boldmath$ P_{11}$ partial wave} is largely non-resonant.
Two $ P_{11}$ resonances were needed to describe this partial wave,
the Roper  resonance and a second one situated in the region
1.84-1.89 GeV/c$^2$. Detailed information on the $ P_{11}$-partial
wave is given in an accompanying letter~\cite{Sarantsev:2007}.

The reaction $\gamma p\to p\pi^0\pi^0$ gives access to the isobar
decomposition of proton-plus-two-pion decays of baryon resonances.
The important intermediate states are $\Delta(1232)\pi$, $
N(\pi\pi)_{S}$, \roper$\pi$ and \trma$\pi$ (see Table
\ref{resonance_list}). The $N(\pi\pi)_{\rm S}$-wave contributes
significantly in the $3^{\rm rd}$ resonance region in which the
three states \trmb\,, \fvma\,, and \fvpa\ are shown to have
non-negligible couplings to $N(\pi\pi)_{\rm S}$. The \trmb\ and
$\Delta(1620)S_{31}$ decay with a significant fraction into $
P_{11}(1440)\pi$, a decay mode which has not yet been reported for
these resonances. Naively, this decay mode is expected to be
suppressed by either the orbital angular momentum barrier and/or by
the smallness of the available phase space.

New and unexpected results were obtained for decays into
$\Delta\pi$. The $\Delta\pi$-contribution clearly dominates the
cross section, especially at lower energies (Fig.\ref{gg_vs_gg}). An
interesting pattern of partial decays of resonances into $\Delta\pi$
is observed which is neither expected by phase space arguments nor
by quark model calculations. $ D_{13}$-decays into
$\Delta\pi_{(S-wave)}$ are allowed by all selection rules but are
observed to be weaker than naively expected. The \trma\ decays into
$\Delta\pi$ in D--wave with about the same strength as in S-wave
even though the orbital angular momentum barrier should suppress
D-wave decays for such small momenta ($\sim 250$\,MeV/c). The \trmb\
$\Delta\pi_{(S-wave)}$-decay is observed to be weaker than
$\Delta\pi_{(D-wave)}$. For both $ D_{13}$-states, the
$\Delta\pi_{(S-wave)}$ seems to be suppressed dynamically. For other
resonances, like \fvma, and \fvpa\, the lower orbital momentum
partial wave is preferred.  The \srma\ and \srmb\ resonances show
sizable couplings to $\Delta\pi$, even though $L=2$ is required. The
\dtma\ state decays dominantly into $\Delta\pi$. Unfortunately no
statement on the dominance of the S- or D-wave decay of the \dtma\
can be made. Two distinct solutions have been found; for one of them
the S-wave, for another one the D-wave, dominates clearly. The
forthcoming double polarisation experiments will help to resolve
this ambiguity.

The results on the decays can be compared to model calculations by
Capstick and Roberts (A); Koniuk and Isgur (B); Stassart and Stancu
(C); Bijker, Le Yaouanc, Oliver, P\`ene and Raynal (D), and Iachello
and Leviatan (E); (numbers and references can be found in
\cite{Capstick:2000qj}, Table VI and VII). A quality factor (mean
fractional deviation) can be defined by the fractional difference
between prediction $x_i$ and experimental result $y_i$ as {
$q_i=4(|x_i|-|y_i|)^2/(|x_i|+|y_i|)^2$}. The $x_i,y_i$ are
proportional to the amplitude for a decay, they are normalized to
give $x_i^2=\Gamma_i$. The $x_i$ carry a signature which is not
given for all calculations. To enable a meaningful comparison, only
absolute values are considered in the comparison. The rms value of
the 14 $q_i$ values is calculated for each model to define a `model'
quality.

 \begin{equation} \scriptsize
q_A=0.271; \quad \scriptsize q_B=0.247; \quad \scriptsize
q_C=0.328; \quad \scriptsize q_D=0.222; \quad \scriptsize
q_E=0.219\,. \label{comparison} \end{equation}  The model
(A) is the only model which predicts the
correct signature in 13 out of the 14 cases. This achievement is not
taken into account in the comparison (\ref{comparison}). The (formally)
most successful model describes baryons in terms of rotations and
vibrations of strings and their algebraic relations
\cite{Bijker:1996tr}.

Summarizing, we have presented new data on the reaction $\gamma p\to
p\pi^0\pi^0$. The partial wave analysis reveals various contributions
to the 2$^{nd}$ and 3$^{rd}$ resonance region. Most masses and widths
determined here are in reasonable agreement with known resonances. Yet,
several $ p2\pi^0$-decay widths contradict expectation. An
interesting pattern of partial decays of resonances into $\Delta\pi$
is observed which was not predicted by quark model calculations.
Several $ p2\pi$-partial widths for baryon resonances in the
2$^{nd}$ and 3$^{rd}$ resonance region and the excitation functions for
$\gamma p \to \Delta\pi$ and $ \gamma p \to N(\pi\pi)_{s}$ have
been determined for the first time.

We would like to thank the technical staff of the ELSA machine
groups and of all the participating institutions of their invaluable
contributions to the success of the experiment. We acknowledge
financial support from the Deutsche Forschungsgemeinschaft (DFG)
within the SFB/TR16 and from the Schweizerische Nationalfond. The
collaboration with St. Petersburg received funds from DFG and RFBR.
U.Thoma thanks for an Emmy Noether grant from the DFG.
A.V.~Sarantsev acknowledges support from RSSF. This work comprises
part of the thesis of M.~Fuchs.

\end{document}